\documentclass[12pt,preprint]{aastex}
\shorttitle{Cold Cloud Inside Local Bubble}
\shortauthors{Meyer et al.\ 2006}
\begin{document}


\title{A Cold Nearby Cloud Inside the Local Bubble}


\author{David M. Meyer\altaffilmark{1} and J. T. Lauroesch\altaffilmark{2}}
\affil{Department of Physics and Astronomy, Northwestern University,
        Evanston, IL  60208}
\email{davemeyer@northwestern.edu, jtl@elvis.astro.northwestern.edu}

\and

\author{Carl Heiles, J. E. G. Peek, and Kyle Engelhorn}
\affil{Astronomy Department, University of California at Berkeley,
        601 Campbell Hall, Berkeley, CA  94720-3411}
\email{heiles@astro.berkeley.edu, goldston@astro.berkeley.edu,
        kengelho@ugastro.berkeley.edu}


\altaffiltext{1}{Visiting Astronomer, Kitt Peak National Observatory,
National Optical Astronomy Observatory, which is operated by the
Association of Universities for Research in Astronomy, Inc. under
cooperative agreement with the National Science Foundation.}
\altaffiltext{2}{Current Address: Department of Physics and Astronomy,
University of Louisville, Louisville, KY  40292}


\begin{abstract}
The high-latitude Galactic H~I cloud toward the extragalactic radio
source 3C~225 is characterized by very narrow 21~cm emission and
absorption indicative of a very low H~I spin temperature of about 20~K.
Through high-resolution optical spectroscopy, we report the detection
of strong, very narrow Na~I absorption corresponding to this
cloud toward a number of nearby stars.  Assuming that the turbulent
H~I and Na~I motions are similar, we derive a cloud temperature of
20$^{+6}_{-8}$~K (in complete agreement with the 21~cm
results) and a line-of-sight turbulent velocity of
0.37~$\pm$~0.08~~km~s$^{-1}$ from a comparison of the
H~I and Na~I absorption linewidths.  We also place a firm upper
limit of 45~pc on the distance of the cloud, which situates it well
inside the Local Bubble in this direction and makes it the
nearest-known cold diffuse cloud discovered to date.
\end{abstract}


\keywords{ISM: atoms -- ISM: clouds -- ISM: structure -- Galaxy: solar
           neighborhood}



\section{Introduction}

Recent H~I 21~cm absorption surveys have revealed a significant population
of cold (T~$<$~40~K) diffuse clouds in the Galactic interstellar medium (ISM)
\citep{hei03,gib05,kav05}.  Two of the most remarkable examples of such
clouds were actually discovered long ago by \citet{ver69} through their very
narrow 21~cm emission.  He found two closely-spaced, degree-sized patches
of H~I gas at $l\sim226\arcdeg,b\sim+44\arcdeg$ (``Cloud A'') and
$l\sim236\arcdeg,b\sim+46\arcdeg$ (``Cloud B'') each with line widths
indicative of kinetic temperatures below 30~K.  \citet{kna72} subsequently
mapped the 21~cm emission of these clouds at higher resolution and
derived respective H~I spin temperatures of $\approx$24~K and $\approx$17~K
for Cloud~A and Cloud~B.  In another approach, \citet{cro85} measured spin
temperatures of $\approx$20~K amd $\approx$14~K from the Cloud~A and Cloud~B
21~cm absorption toward the extragalactic radio sources 3C~225 and 3C~237.

As part of their Millenium Arecibo 21~cm Absorption-Line Survey,
\citet{hei03} revisited these two cold clouds in the context of new
observations of 3C~225 and 3C~237.  Utilizing the Leiden-Dwingeloo 21~cm
sky survey data \citep{har97}, they found that Cloud~A and Cloud~B are the
predominant components of a narrow ($\approx$2$\arcdeg$), broken ribbon of
cold H~I gas stretching over 20$\arcdeg$ across the constellation Leo.  Their
observations of 3C~225a,b (6.3$\arcmin$~separation) yield spin temperatures
of 22~K and 17~K and H~I column densities of $1.7\times10^{19}$~cm$^{-2}$
and $3.2\times10^{19}$~cm$^{-2}$ for the intervening Cloud~A gas.
Assuming thermal pressure equilibrium under these conditions leads to an
extremely thin, sheetlike geometry for the cloud \citep{hei03}.  A more
detailed physical assessment including the specific aspect ratio, actual
size, and mass of the cloud is limited by the unknown cloud distance.
In this {\it Letter}, we present the first optical absorption-line study 
of the cold Leo clouds and conclusively show that Cloud~A
is located well inside the Local Bubble.

\section{Observations}

Observations of the interstellar Na~I~D$_2$ $\lambda$5889.951
and D$_1$ $\lambda$5895.924
absorption toward 33 stars in Leo were obtained in 2006~January
and March using the 0.9~m Coude Feed telescope and spectrograph at Kitt
Peak National Observatory.  The spectrograph was configured with the long
focal length camera (Camera~6), the echelle grating, a cross-dispersing
grism (\#770), and the T2KB Tek2048 (2006~January) and
T1KA Tek1024 (2006~March) CCD detectors.
The 150~$\mu$m slit width resulted in a 2.1~pixel spectral resolution of
0.026~\AA\ or 1.3~km~s$^{-1}$ at the Na~I wavelengths as measured from
the Th-Ar calibration lamp spectra.  The NOAO IRAF echelle data reduction
package was utilized to bias-correct, scattered-light-correct, flat-field,
wavelength-calibrate, order-extract, and sum the individual CCD exposures
into the final net spectra of each object.  Observations of the
nearby star $\alpha$~Leo were also obtained and reduced in a similar
manner in order to create an essentially ISM-free template to divide out
the telluric absorption in the vicinity of the Na~I lines toward the
program stars.

Figure~1 illustrates a sample of our interstellar Na~I D$_2$ spectra
toward 5 stars in the sky region of 3C~225.  They all clearly show a
strong, very narrow absorption component at the same LSR velocity
($+$4~km~s$^{-1}$) as the narrow H~I 21~cm absorption observed toward
3C~225.  At our spectral resolution, this Na~I component is so narrow
that it can be measured without significant stellar line blending
uncertainties toward stars with spectral types as late as M2~III
(except in those cases where the velocity of the component is close to
that of the strong stellar Na~I absorption in these later-type stars).
This capability dramatically expands the sky coverage and distance
range of suitable stellar targets well beyond the sparse sample of
early-type stars that are traditionally employed for interstellar
absorption-line studies.  Our program sample of 33~stars ranges in
spectral type from B9~V to M2~III and represents most of the bright
targets accessible with a small telescope in the sky vicinity of
the cold Leo clouds.

We detect Na~I absorption at the velocity corresponding to the cold
H~I gas in 23 of our targeted Leo sightlines.  Utilizing the FITS6P
profile-fitting package \citep{wel94} and taking into account the
hyperfine splitting of the Na~I~D lines ($\Delta$v~$=$~1.05~km~s$^{-1}$),
we find that a single-component solution to both the D$_2$ and D$_1$
profiles provides an excellent fit to this absorption in all cases.
Figure~2 illustrates the single-component fit (and the intrinsic
hyperfine-split profile) for the D lines toward HD~83509 which
exhibits the narrowest Na~I linewidth (b~$=$~0.33~km~s$^{-1}$) among
our measurements of the cold gas absorption.
Table~1 lists the Na~I column
densities, Gaussian linewidth parameters (b-values), and LSR velocities
measured for each detection of this absorption and 2$\sigma$ upper
limits on N(Na~I) for each non-detection.  As is the case with the
corresponding H~I 21~cm emission \citep{hei03}, the LSR velocity of this
Na~I component generally decreases from about $+$4 to $+$2~km~s$^{-1}$
with increasing Galactic longitude across Leo.

\section{Discussion}

Figure~3 illustrates the strength of the Na~I absorption at
v$_{LSR}$~$\approx$~3~km~s$^{-1}$ toward our program stars as a
function of sky position relative to the 21~cm emission contour of the
cold H~I gas \citep{hei03} and the location of 3C~225 and 3C~237.
It is clear from Figure~3 that the Na~I absorption is an excellent
tracer of the H~I emission for this cold gas.  In the case of
comparisons with the 21~cm absorption observed toward 3C~225a,b,
four stars with matching-velocity Na~I absorption (HD~84194,
HD~84182, HD~83683, and HD~83509) are located within 1$\arcdeg$ of
this radio source.  The mean linewidth of this Na~I absorption (and the
standard error of the mean) is b~$=$~0.54~$\pm$~0.10~km~s$^{-1}$
(0.90~$\pm$~0.17~km~s$^{-1}$~FWHM) as compared with
the 1.3~km~s$^{-1}$~FWHM H~I linewidth measured toward both
3C~225a and 3C~225b by \citet{hei03}.  Assuming that the turbulent
H~I and Na~I motions are similar, a comparison of the H~I and
Na~I linewidths through the expression $b^2=(2kT/m)+2v_t^2$
yields a cloud kinetic temperature of 20$^{+6}_{-8}$~K
and a one-dimensional rms turbulent
velocity (v$_t$) of 0.37~$\pm$~0.08~km~s$^{-1}$.  Thus, the
Na~I measurements are completely consistent with the remarkably
cold temperatures previously derived for Cloud~A from the H~I
21~cm spin excitation toward 3C~225.

The Na~I observations also provide a stringent constraint on the
distance of Cloud~A.  The two closest program stars within the
Cloud~A H~I contour are at {\it Hipparcos} distances \citep{per97}
of 42.1$^{+1.8}_{-1.6}$~pc (HD~83683) and 42.8$^{+1.9}_{-1.7}$~pc
(HD~85091) and they both exhibit strong Cloud~A Na~I absorption.
Another nearby star at a distance of 41.5$^{+1.7}_{-1.6}$~pc
(HD~83808) just outside the Cloud~A contour exhibits very weak
Na~I absorption at the appropriate velocity.  The only program
star closer than these objects is at a distance of
39.2$^{+1.5}_{-1.4}$~pc (HD~80218) and it lies just outside the
H~I contour of the lowest-longitude cold Leo cloud.  Consequently,
the lack of Na~I absorption toward HD~80218 is not conclusive
with respect to the distance of this cloud.  Although it is likely
that the three cold Leo clouds illustrated in Figure~3 are at the
same distance based on their similar velocities, narrow linewidths,
and close position on the sky, the Na~I measurements are not
definitive in this regard.  Nevertheless, the
Na~I absorption observed toward HD~83683, HD~85091, and
HD~83808 does allow us to place a firm upper limit of 45~pc on the
distance of Cloud~A.  With this limit, Cloud~A is not only one of the
coldest diffuse clouds discovered to date, it also becomes the
nearest-known cold interstellar cloud to the Sun.

The proximity of Cloud~A is especially interesting in that it places
this cold cloud well inside the Local Bubble of hot ($\sim$10$^6$~K),
tenuous (n$_H$~$\sim$~0.01~cm$^{-3}$) gas surrounding the Sun out to
distances of $\sim$100~pc or more \citep{cox87}.  The Local Bubble is
believed to have originated through multiple supernovae explosions
over the past $\sim$5 to $\sim$15~million years \citep{smi01,bre06}.
Utilizing Na~I absorption measurements toward $\approx$1000
early-type (earlier than A5) stars with {\it Hipparcos} distances,
\citet{lal03} have mapped the neutral gas ``edge'' of the Local
Bubble across the sky.  In the direction of the cold Leo clouds,
they place this boundary at a distance of about 100~pc.  Since
our star sample in the Leo field of Figure~3 is much larger (33)
than that (2) of Lallement et al.\ (due in part to our inclusion of
objects with later spectral types), we can explore this boundary
in greater detail.  Among the 17~program stars with d~$>$~157~pc,
nine exhibit interstellar Na~I absorption at LSR velocities (ranging
from $-$11.5 to $-$1.6~km~s$^{-1}$) distinct from those of the cold
Leo clouds whereas none of the 16 stars with d~$<$~157~pc display such
absorption.  As illustrated in Figure~1, such distance distinctions
are also apparent when the Na~I spectra are compared in localized
sky regions.  Allowing for patchiness in the more distant Na~I
absorption and the larger stellar distance uncertainties past
$\approx$100~pc, the nearest neutral gas clouds beyond Cloud~A are
at a distance between 100 and 150~pc.  Thus, Cloud~A
is indeed closer to the Sun than it is to the neutral gas
boundary of the Local Bubble.

Within the Local Bubble, there are known to be a number of warm,
partially-ionized clouds which collectively have an average
temperature of 6700~K and a mean thermal pressure (P/k) of
2300~K~cm$^{-3}$ \citep{red04}.  Although the thermal pressures of
these clouds appear to be below that of the surrounding hot gas in
the Local Bubble, they provide a benchmark in considering the
physical characteristics of Cloud~A.  Assuming that Cloud~A is in
thermal pressure equilibrium at 2300~K~cm$^{-3}$, the 20~K
temperature and mean N(H~I) of $2.5\times10^{19}$~cm$^{-2}$ toward
3C~225 yield a cloud thickness of $2.2\times10^{17}$~cm or
0.07~pc (this equilibrium thickness would be $\approx$4~times less
at the hot gas thermal pressure).
If Cloud~A is at a distance of 40~pc, its
$2\arcdeg\times7\arcdeg$ H~I angular extent would then correspond to
a physical size of $1.4\times4.9$~pc, a very thin
length-to-width-to-thickness aspect ratio of 70:20:1,
and an H~I mass of about 1.4~solar masses.  At
the other extreme, if we assume that Cloud~A is at a distance
($\approx$2~pc) where its projected width is equal to its
equilibrium thickness of 0.07~pc, the cloud size would be
$0.07\times0.25$~pc with an H~I mass of about 3.6~Jupiter masses.
In any case, given our measured v$_t$ of 0.37~km~s$^{-1}$, the
line-of-sight turbulent crossing time over the equilibrium
thickness of 0.07~pc is $1.8\times10^5$~years.  In other words,
it appears that Cloud~A is a much more transient phenomenon than
the Local Bubble and thus, is likely to have originated somehow
in this seemingly hostile environment to cold clouds.

Recently, several theoretical studies have proposed the formation
of cold clouds with filamentary or sheet-like geometry through
converging flows of warm interstellar gas \citep{aud05,vaz06}.
In particular, Vazquez-Semadeni et al.\ find that they can come
close to reproducing the \citet{hei03} description of Cloud~A
through transonic compression in colliding warm gas streams.
They acknowledge that their timescale to produce this cold structure
is rather long ($\sim$1~Myr) but may be consistent with the observations
if Cloud~A is overpressured and v$_t$ reflects the gas flow rather
than turbulent motions.  Such models are promising but will need to
consider other factors such as magnetic fields and certainly thermal
conduction in order to provide a full explanation of the nature
and evolution of Cloud~A inside the hot Local Bubble.  For example,
\citet{sla06} has shown that the evaporative and condensation time
scales for such cold clouds are measured in millions of years
even if immersed in a hot surrounding medium.  It is also possible
that the hot gas in the Local Bubble may be less pervasive than
generally thought, especially given recent results suggesting 
significant foreground contamination by heliospheric 
soft X-ray emission \citep{lal04}.  In any case,
the size of Cloud~A on the sky and its isolated nature make it ideal for
further optical/UV absorption-line and H~I 21~cm emission-line studies
of its distance, small-scale structure, velocity fields,
and other physical characteristics such as
dust depletion, electron density, and radiation field.
A better understanding of the origin and survival of this nearby 20~K
cloud inside the Local Bubble will almost certainly have a bearing on
interpreting the continuing discoveries of such cold diffuse
clouds elsewhere in the Galactic ISM.



\acknowledgments

It is a pleasure to thank Daryl Willmarth for his assistance with the
KPNO observations (which included de-icing the Coude Feed telescope
enclosure after an immobilizing March blizzard).  We are also grateful
to the anonymous referee whose helpful comments improved the paper.
C. H. acknowledges support from NSF grant AST 04-06987.



{\it Facilities:} \facility{KPNO (Coude Feed)}





\clearpage

\begin{deluxetable}{crrcc}
\tabletypesize{\scriptsize}
\tablecolumns{5}
\tablecaption{Na I Measurements of the Cold Leo Gas\label{tbl-1}}
\tablewidth{0pt}
\tablehead{
\colhead{Star\tablenotemark{a}} & \colhead{Distance\tablenotemark{b}} &
\colhead{N(Na I)\tablenotemark{c}} & \colhead{b(Na I)\tablenotemark{d}} &
\colhead{v(Na I)\tablenotemark{e}} \\
\colhead{} & \colhead{(pc)} & \colhead{(cm$^{-2}$)} &
\colhead{(km s$^{-1}$)} & \colhead{(km s$^{-1}$)}}
\startdata
HD 80218 & 39.2$^{+1.5}_{-1.4}$ & $<5.0\times10^9$ & \nodata & \nodata \\
HD 80510 & 174$^{+33}_{-24}$ & $4.20\times10^{11}$ & 0.39 & $+4.48$ \\
HD 80652 & 128$^{+16}_{-14}$ & $2.04\times10^{11}$ & 0.56 & $+4.32$ \\
HD 80613 & 233$^{+58}_{-39}$ & $<5.9\times10^9$ & \nodata & \nodata \\
HD 83023 & 156$^{+30}_{-21}$ & $2.24\times10^{11}$ & 0.84 & $+4.34$ \\
HD 82906 & 254$^{+187}_{-76}$ & $3.23\times10^{11}$ & 0.54 & $+4.15$ \\
HD 83343 & 64.5$^{+5.7}_{-4.9}$ & $2.08\times10^{11}$ & 0.58 & $+4.08$ \\
HD 84194 & 219$^{+70}_{-43}$ & $5.00\times10^{11}$ & 0.60 & $+4.20$ \\
HD 83509 & 60.5$^{+3.9}_{-3.5}$ & $4.55\times10^{11}$ & 0.33 & $+3.99$ \\
HD 84182 & 158$^{+38}_{-25}$ & $1.29\times10^{12}$ & 0.78 & $+3.79$ \\
HD 83683 & 42.1$^{+1.8}_{-1.6}$ & $5.05\times10^{11}$ & 0.43 & $+4.01$ \\
HD 83362 & 304$^{+159}_{-78}$ & $<8.9\times10^9$ & \nodata & \nodata \\
HD 84937 & 80.4$^{+7.5}_{-6.3}$ & $1.62\times10^{11}$ & 0.92 & $+2.94$ \\
HD 82395 & 73.2$^{+4.9}_{-4.4}$ & $<4.7\times10^{9}$ & \nodata & \nodata \\
HD 85268 & 568$^{+568}_{-190}$ & $2.01\times10^{11}$ & 1.17 & $+2.92$ \\
HD 84561 & 215$^{+53}_{-35}$ & $8.20\times10^{11}$ & 0.64 & $+3.86$ \\
HD 84722 & 89.8$^{+8.8}_{-7.4}$ & $7.70\times10^{11}$ & 0.64 & $+3.69$ \\
HD 85259 & 275$^{+82}_{-51}$ & $3.75\times10^{11}$ & 0.40 & $+4.13$ \\
HD 86360 & 162$^{+25}_{-18}$ & $4.19\times10^{10}$ & 0.63 & $+1.89$ \\
HD 84183 & 76.6$^{+5.8}_{-5.0}$ & $<9.4\times10^{9}$ & \nodata & \nodata \\
HD 85091 & 42.8$^{+1.9}_{-1.7}$ & $8.58\times10^{11}$ & 0.62 & $+3.47$ \\
HD 83808 & 41.5$^{+1.7}_{-1.6}$ & $3.31\times10^{10}$ & 2.07 & $+4.40$ \\
HD 85269 & 196$^{+47}_{-31}$ & $5.43\times10^{11}$ & 0.60 & $+3.80$ \\
HD 84764 & 247$^{+104}_{-57}$ & $<4.0\times10^{10}$ & \nodata & \nodata \\
HD 86080 & 207$^{+39}_{-29}$ & $<2.6\times10^{10}$ & \nodata & \nodata \\
HD 87837 & 84.1$^{+5.4}_{-4.8}$ & $<2.4\times10^{10}$ & \nodata & \nodata \\
HD 86663 & 161$^{+23}_{-18}$ & $<8.6\times10^9$ & \nodata & \nodata \\
HD 83425 & 84.0$^{+7.3}_{-6.2}$ & $<5.3\times10^9$ & \nodata & \nodata \\
HD 88048 & 191$^{+37}_{-27}$ & $5.79\times10^{11}$ & 0.57 & $+2.36$ \\
HD 87682 & 106$^{+11}_{-9}$ & $3.73\times10^{11}$ & 0.62 & $+3.24$ \\
HD 87392 & 126$^{+21}_{-16}$ & $4.60\times10^{11}$ & 0.71 & $+2.96$ \\
HD 88315 & 251$^{+81}_{-50}$ & $9.15\times10^{11}$ & 1.01 & $+2.85$ \\
HD 88547 & 162$^{+27}_{-21}$ & $5.71\times10^{11}$ & 0.60 & $+2.32$ \\
\enddata
\tablenotetext{a}{The program stars ordered by increasing Galactic
longitude.}
\tablenotetext{b}{The stellar distances and their $\pm$1$\sigma$ uncertainties
as derived from {\it Hipparcos} parallax measurements \citep{per97}.}
\tablenotetext{c}{The 1$\sigma$ errors in the measured Na I column
densities are all less than 6\% of the listed N(Na I) values.  The stars
with no cold Leo Na I absorption are listed with a 2$\sigma$ N(Na I)
upper limit.}
\tablenotetext{d}{The 1$\sigma$ uncertainties in the measured Gaussian line
widths are all less than 8\% of the listed b(Na I) values.}
\tablenotetext{e}{The 1$\sigma$ uncertainties in the measured velocities of
the cold Leo Na I absorption are all less than 0.08 km s$^{-1}$.}
\end{deluxetable}
\clearpage



\begin{figure}
\epsscale{0.7}
\plotone{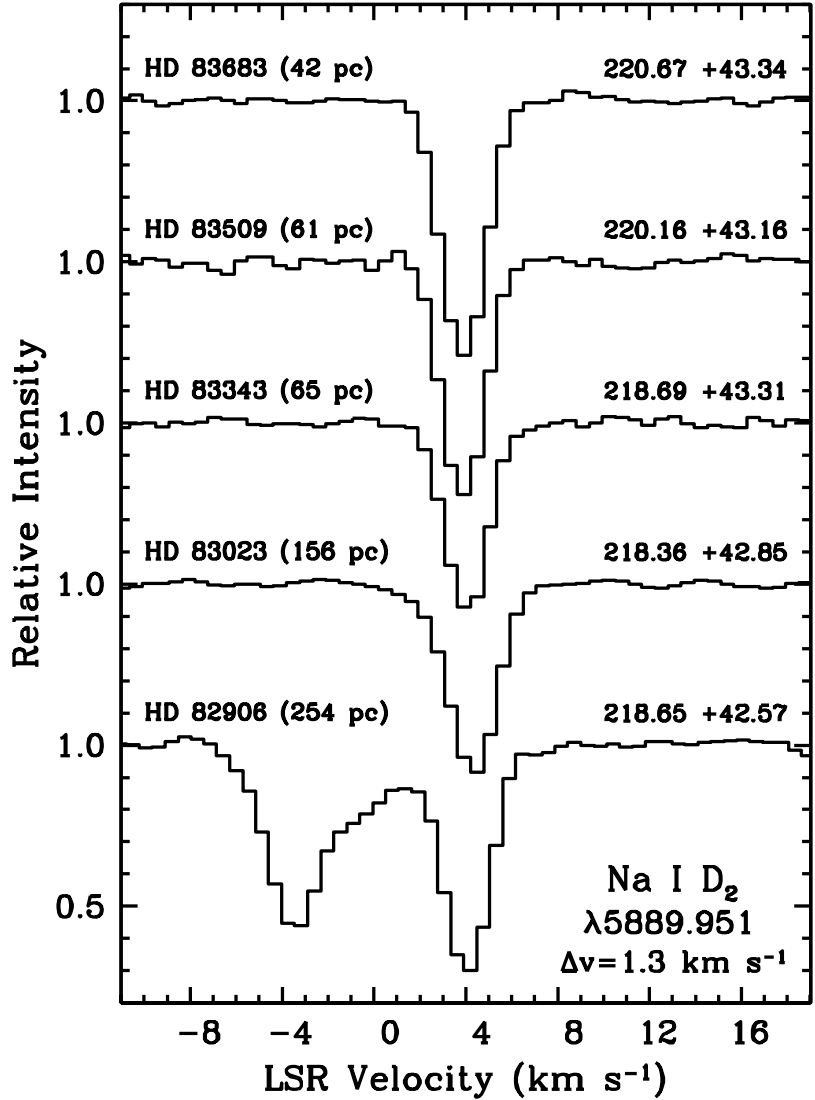}
\caption{High-resolution ($\Delta$v~$=$~1.3~km~s$^{-1}$) KPNO Coude Feed
spectra of the interstellar Na~I~D$_2$ absorption toward a sample of
stars in the sky vicinity of 3C~225.  Arranged in descending order of
increasing stellar distance, all of these spectra exhibit a strong, very
narrow Na~I component at v$_{LSR}$~$\approx$~4~km~s$^{-1}$ corresponding
to Cloud~A.  The most distant star (HD~82906) exhibits additional Na~I
absorption at lower LSR velocities that is presumably due to neutral
gas near the Local Bubble boundary.\label{fig1}}
\end{figure}

\clearpage


\begin{figure}
\epsscale{0.7}
\plotone{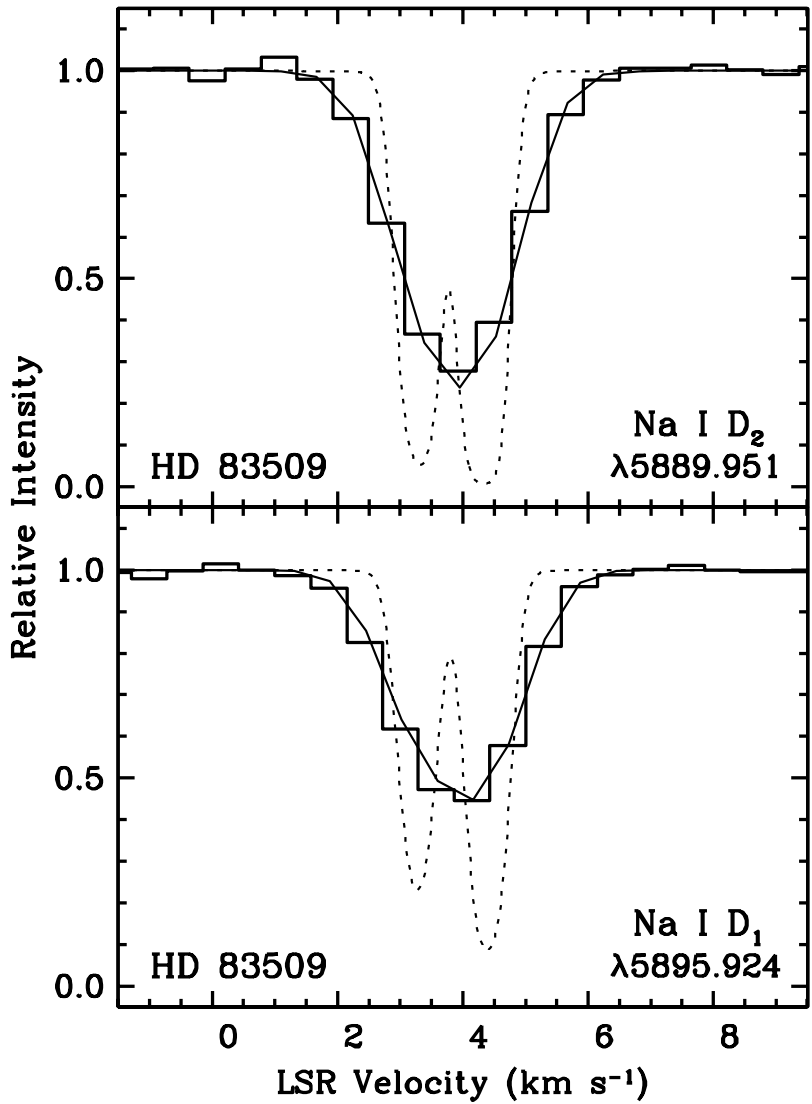}
\caption{The interstellar Na~I D$_2$ $\lambda$5889.951 (top) and D$_1$
$\lambda$5895.924 (bottom) absorption profiles toward HD~83509.  The smooth
lines through the data represent the best single-velocity-component
Gaussian fit to both of the observed profiles (the oscillator strength
of the D$_2$ line is twice that of the D$_1$ line).
This fit is characterized
by an intrinsic hyperfine-split ($\Delta$v~$=$~1.05~km~s$^{-1}$)
Na~I~D profile with a linewidth of b~$=$~0.33~km~s$^{-1}$ (denoted by the
dotted lines) broadened by the instrumental resolution of
1.3~km~s$^{-1}$.  The Na~I linewidth toward HD~83509 is the narrowest
measured among the 23 stars exhibiting cold Leo gas absorption.
\label{fig2}}
\end{figure}

\clearpage

\begin{figure}
\epsscale{1.0}
\plotone{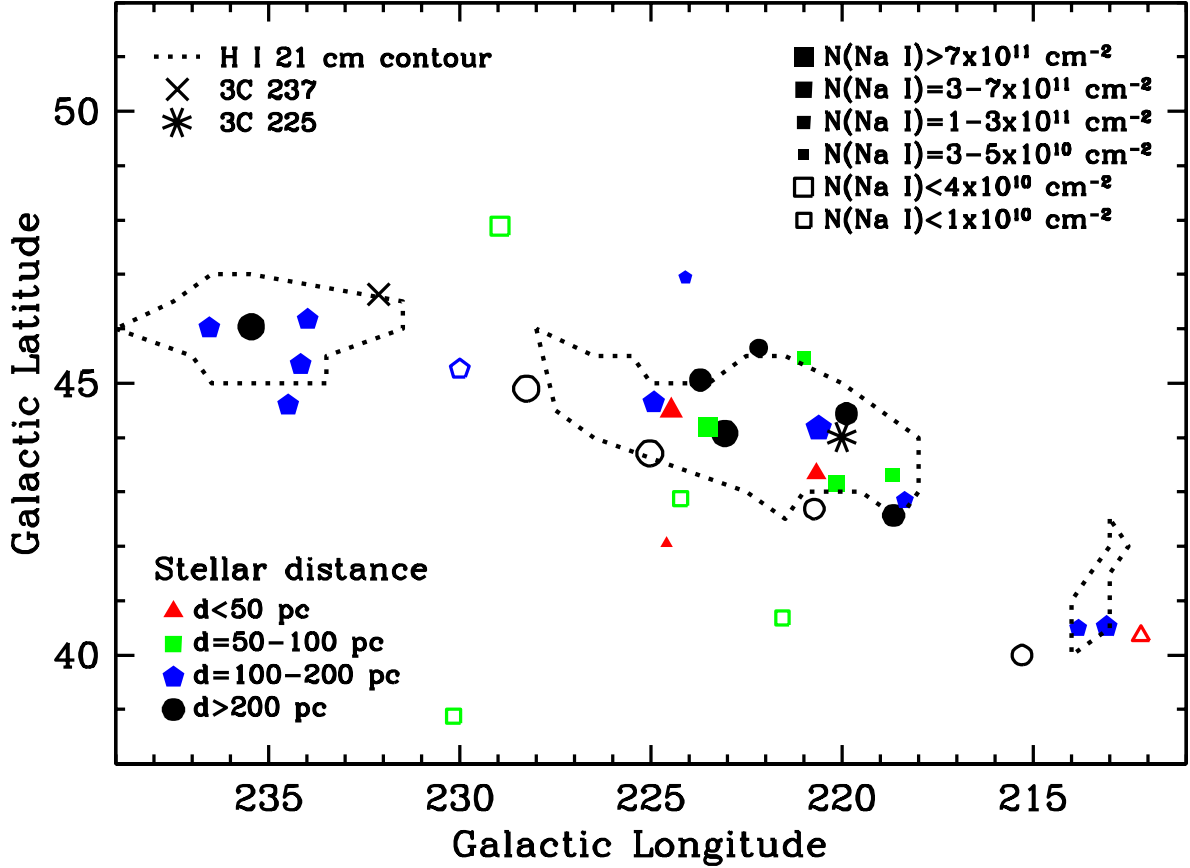}
\caption{The Na~I column densities of the cold Leo gas as a function
of stellar distance and sky position relative to the H~I 21~cm contours
of \citet{hei03}.  In order to present the appropriate aspect ratios
for the clouds denoted by these contours, the horizontal length scaling
is 71\% (cos~45$\arcdeg$) of the vertical scaling to account for the
longitude foreshortening at higher latitudes.
The filled symbols represent Na~I detections of the cold gas,
the open symbols indicate non-detections, and the symbol sizes
reflect the measured Na~I column densities (or upper limits) as denoted
in the upper-right symbol legend.  The symbol shapes (color-coded in
the online version) reflect the stellar distances as shown in the
lower-left symbol legend.  The sky positions of the extragalactic radio
sources 3C~225 (within the Cloud~A H~I contour) and
3C~237 (on the Cloud~B H~I contour) are also
displayed for reference purposes.\label{fig3}}
\end{figure}





\end{document}